\documentclass{article}       % onecolumn (second format)
\usepackage{color}
\usepackage{graphicx}% Include figure files
\usepackage{dcolumn}% Align table columns on decimal point
\usepackage{bm}% bold math
\usepackage{physics}
\usepackage{sidecap}
\usepackage{blindtext}
\usepackage{float}
\usepackage{indentfirst}
 \usepackage{amsthm}
\usepackage{amssymb}
\usepackage{graphicx,wrapfig}
\newcommand*\diff{\mathop{}\!\mathrm{d}}

\usepackage{algorithm}
\usepackage[noend]{algpseudocode}
\newcommand{\equ}[1]{Eq.~(\ref{#1})}
\newcommand{\fig}[1]{Fig.~\ref{#1}}
\usepackage{amsthm}
\newtheorem*{definition}{Definition}
\newcommand{\secti}[1]{Sec.~\ref{#1}}
 \usepackage{geometry}
 \usepackage{authblk}
\begin{document}

%\title{Spontaneous Emission and Transport Characteristics of a Two-level System Coupled to Multiple Waveguides }
\title{The Effective Geometry Monte Carlo Algorithm: Applications to Molecular Communication }

     % if too long for running head

\author[1,2]{Fatih Dinc}
\affil[1]{ Electrical \& Electronics Engineering Department, Bo\u{g}azi\c{c}i University, Istanbul, 34342, Turkey}
\affil[2]{Perimeter Institute for Theoretical Physics, Waterloo, Ontario, N2L 2Y5, Canada}

\author[2]{Leander Thiele \thanks{The first two authors contributed equally to this work.}}
\author[1]{Bayram Cevdet Akdeniz}

\date{Version: \today} %/ Accepted: date
\maketitle

\begin{abstract}
In this work, we address the systematic biases and random errors stemming from finite step sizes encountered in diffusion simulations. We introduce the Effective Geometry Monte Carlo (EG-MC) simulation algorithm which modifies the geometry of the receiver. We motivate our approach in a 1D toy model and then apply our findings to a spherical absorbing receiver in a 3D unbounded environment. We show that with minimal computational cost the impulse response of this receiver can be precisely simulated using EG-MC. Afterwards, we demonstrate the accuracy of our simulations and give tight constraints on the single free parameter in EG-MC. Finally, we comment on the range of applicability of our results. While we present the EG-MC algorithm for the specific case of molecular diffusion, we believe that analogous methods with effective geometry manipulations can be utilized to approach a variety of problems in other branches of physics such as condensed matter physics and cosmological large scale structure simulations.
\end{abstract}


%\pacs{IEEE}% PACS, the Physics and Astronomy

\section{Introduction}
\label{sec_Introduction}

Understanding the communication between complex systems of diffusing molecules requires fast and reliable simulation methods. While many methods have been suggested in the literature to simulate diffusion processes inside such systems, they mainly focus on changing the path of the diffusing molecules to overcome systematic errors that arise from the finite step size. We first give a brief review of these approaches.

%\subsection{Previous Work}
%\label{subsec_Previous_Work}
Molecular communication is growing in importance \cite{farsad2016comprehensive}. The understanding of communication between nanomachines is extremely important for further development in this area. In this pursuit, molecular communication via diffusion (MCvD) channels has been explored extensively in the literature \cite{yilmaz2014arrival,nakano2012channel}.  Although analytical results are derived for some simple diffusion channels, such as a point transmitter and a spherical absorbing receiver in an unbounded environment \cite{yilmaz2014three} and in bounded environments \cite{dinc2018,al2018modeling}, a deeper analysis including more complicated structures requires fast and reliable simulations of sophisticated diffusion channels. To meet this demand, many different simulation frameworks have been proposed in the literature, such as N3Sim \cite{llatser2014n3sim}, NanoNS \cite{gul2010nanons}, BiNS2 \cite{felicetti2013simulating} and AcCoRD \cite{noel2017simulating}.
The most simple algorithm performing diffusion channel simulations is usually referred to as a Monte Carlo (MC) simulation algorithm. In this algorithm, the molecules iteratively diffuse inside the channel following a Gaussian probability distribution for their step sizes. In each iteration, the molecule position is calculated and molecules that are inside the volume of the receiver are absorbed \cite{yilmaz2014simulation}. This algorithm performs quite accurately for small step sizes while requiring high computation power, i.e. iteration number, in order to simulate complex systems. For large step sizes, the accuracy of the simulation decreases due a combination of different effects. One such effect, called the intra-step absorption problem, is considered in \cite{noel2017simulating,arifler2017monte, wang2018novel}. This effect arises because some molecules close to the receiver are outside of the receiver boundary at two consecutive iteration steps, whereas in reality some portion of them should have been absorbed in-between these steps. In the literature, different absorption models approximating the behaviour of molecules near the receiver boundary (RMC in \cite{arifler2017monte} and APMC in \cite{wang2018novel}) are proposed to mitigate the intra-step absorption problem. In general, these methods perform random variable (RV) calculations to accurately describe the absorption probabilities of the molecules while decreasing the required iteration number. RMC requires that the surface of the receiver can be approximated by a flat surface, whereas APMC remains accurate only for large step sizes. Both methods work quite well in their respective regime of applicability at the expense of requiring an additional path-integral-like RV calculation which adds another layer of complexity to the simulation algorithm. However, intra-step absorption is only one of the problems which arise from the finite step size of MC simulations.

% LFT: I think this is repeating what has been said before
%In general, the main issue with finite step size is the fact that molecules are absorbed at the end of the steps rather than being instantly absorbed upon collusion with the receiver. One way of solving this problem is to change the way the molecules behave around the receivers and this has been the approach of the literature so far.

%\subsection{Our Idea}
%\label{subsec_Our_Idea}
In this work, we take a different perspective on the general problem of finite step size. As a first step, we invoke the principle of locality, which asserts that the state of an object can only be affected through its interactions with its nearby surroundings.
% LFT: I think this is not necessary
%In mathematical terms, if a system has local properties, the dynamics of the system can be uncovered by just considering the small neighborhood of the system.
%The principle suggests an action-at-a-distance scenario and is in the heart of many great advances in physics such as various field theories and relativity. 
While being utilized predominantly in physical contexts, such as field theory and relativity,
the principle of locality is also pivotal to the Effective Geometry Monte Carlo simulation method (EG-MC) we develop in this article, where we exploit the locality of the absorption of molecules by a receiver.
Building up on this intuition, instead of changing the diffusion of molecules near the receiver boundary we introduce an effective absorption surface such that the mean absorption position of the molecules becomes the receiver boundary. By doing so, we both solve the intra-step absorption problem described in the literature \cite{noel2017simulating,arifler2017monte, wang2018novel} and the negative absorption index problem (NAI) which we introduce below. The range of applicability of our method is constrained by the requirement of locality that the step size of the simulation is less than the distance between the receiver and the transmitter.

We organize this paper as follows. We first motivate the negative index problem and introduce the EG-MC algorithm in 1D while showing a way of fixing the free parameter that describes EG-MC. We then apply the EG-MC algorithm to a spherical absorbing receiver described in \cite{yilmaz2014three}. Afterwards, we perform an error analysis to show the consistency of our results in different regimes, where we also illustrate the Poisson noise in the simulation. Finally, we conclude with the range of applicability and a quantitative measure of locality in the absorption process.

\section{Problem Motivation: 1D Toy Model}
\label{sec_problem_motivation}

%\subsection{Definition of the problem}
%\label{subsec_problem_definition}
We shall consider unbounded diffusion channels of arbitrary dimension first. In d-dimensional diffusion channels, the diffusion process is usually modelled as a Brownian motion with a step size
\begin{align}
  &\Delta x_i=\mathcal{N}(0,2D\Delta t), \quad i=1,2,..,d,  \label{brown}
\end{align}
where $D$ is the diffusion coefficient, $\Delta x_i$ ($i=1,2,..,d$) are the incremental step sizes in the d-dimensions, $\Delta t$ is time step, and $ \mathcal{N}(\mu,\sigma^2)$ is the normal distribution with mean $\mu$ and variance $\sigma^2$.

\begin{figure}
    \centering
    \includegraphics[width=5cm]{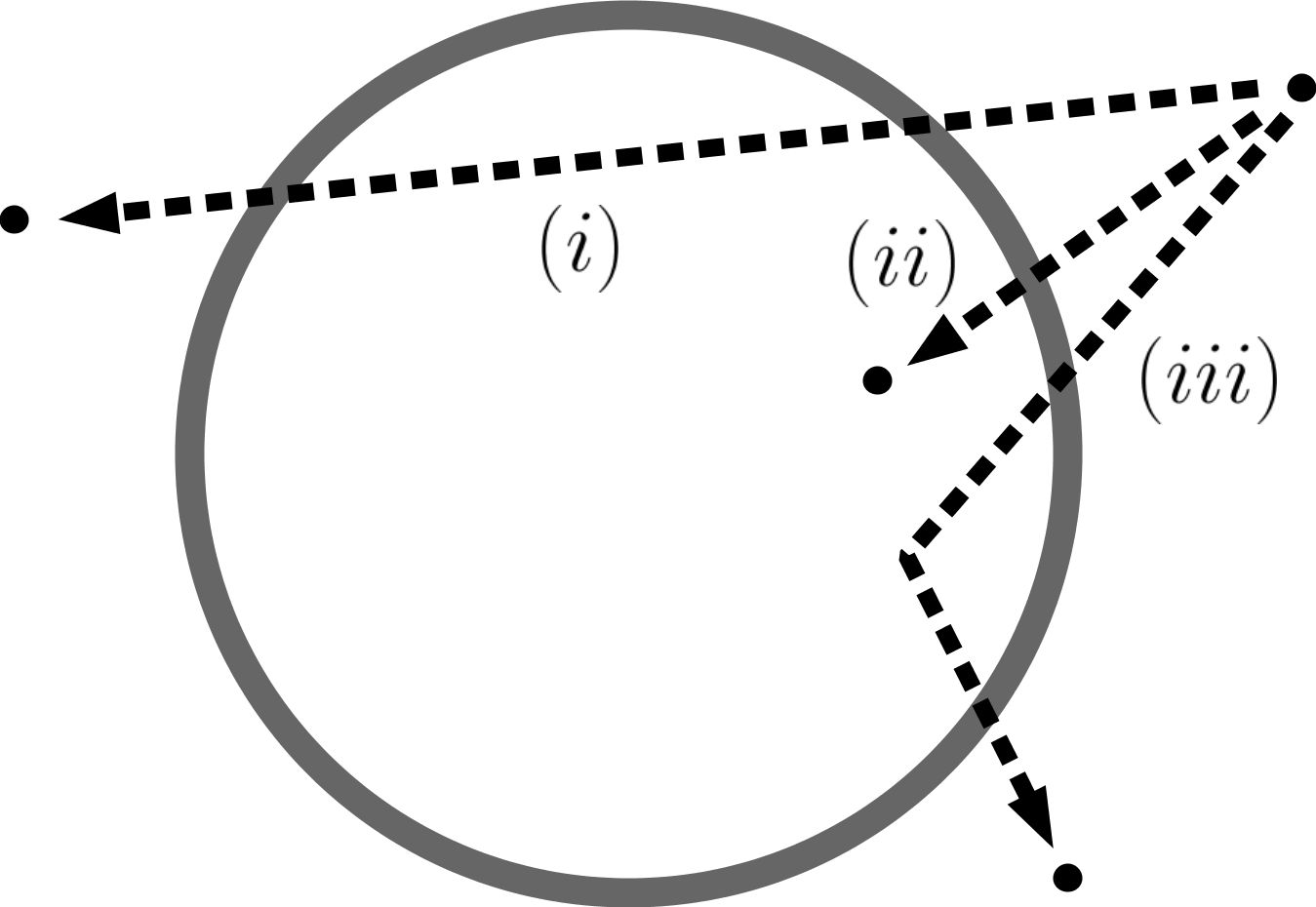}
    \caption{An illustration of various problems resulting from finite step size. (i),(iii): Intra-step absorption problem, (ii): Negative absorption index (NAI) problem} \label{fig:main}
\end{figure}

\begin{figure}
    \centering
    \includegraphics[width=16cm]{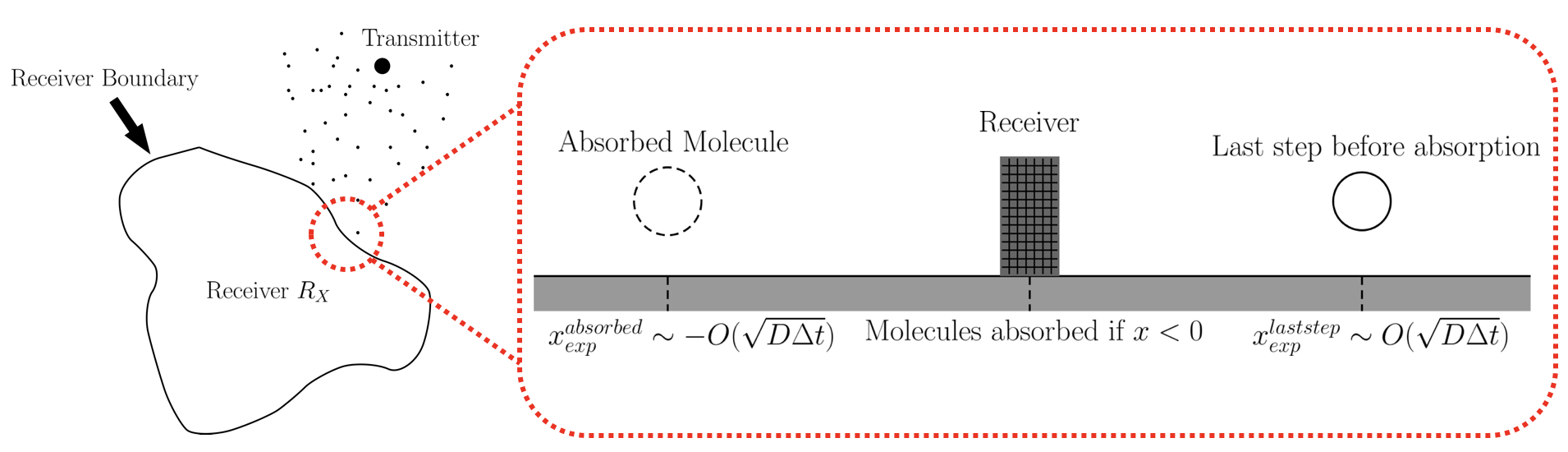}
    \caption{An Illustration of conventional Monte Carlo algorithm for the specific case of 1D. Any smooth surface can be approximated as a 1D as long as the step size is small enough, allowing us to define AI for these surfaces. } \label{fig:MC}
\end{figure}

\begin{figure}
    \centering
     \includegraphics[width=16cm]{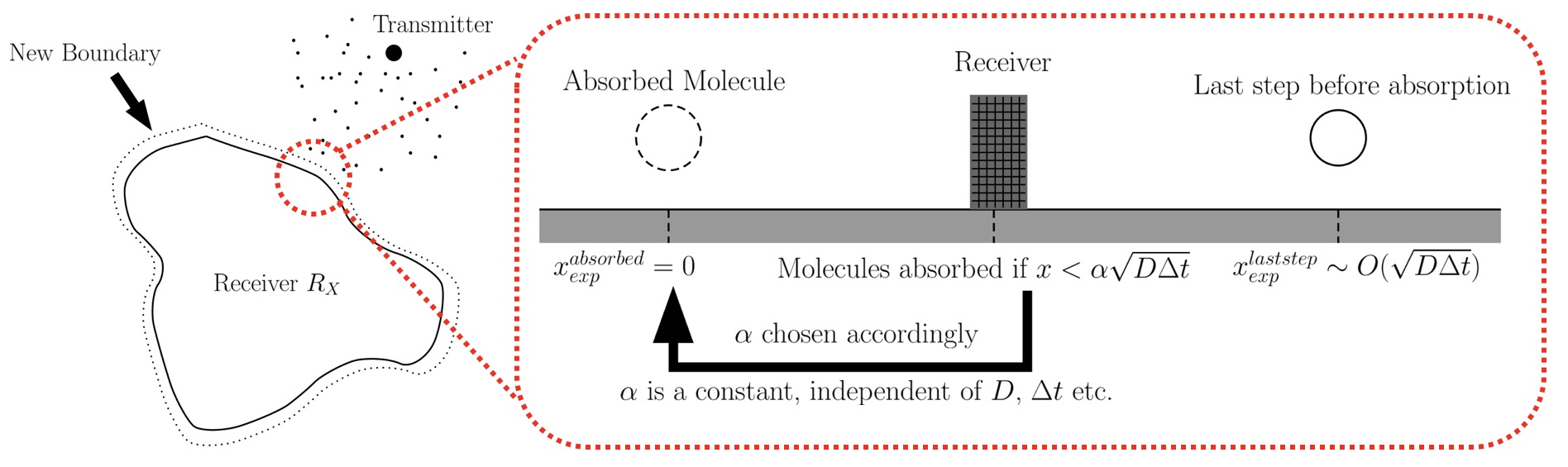}
    \caption{An Illustration of EG-MC Algorithm both in the general 3D case and in the 1D case. A new boundary is created such that the molecules are absorbed right at the boundary by the algorithm. The main motivation behind the shift of the boundary is the principle of locality, which results in linear relation between the free parameter $\alpha$ and the absorption position $x_{\mathrm{exp}}$ (See Fig. \ref{linear}). } \label{fig:EGMC}
\end{figure}

\begin{figure}[!b]
    \centering
    \includegraphics[width=9cm]{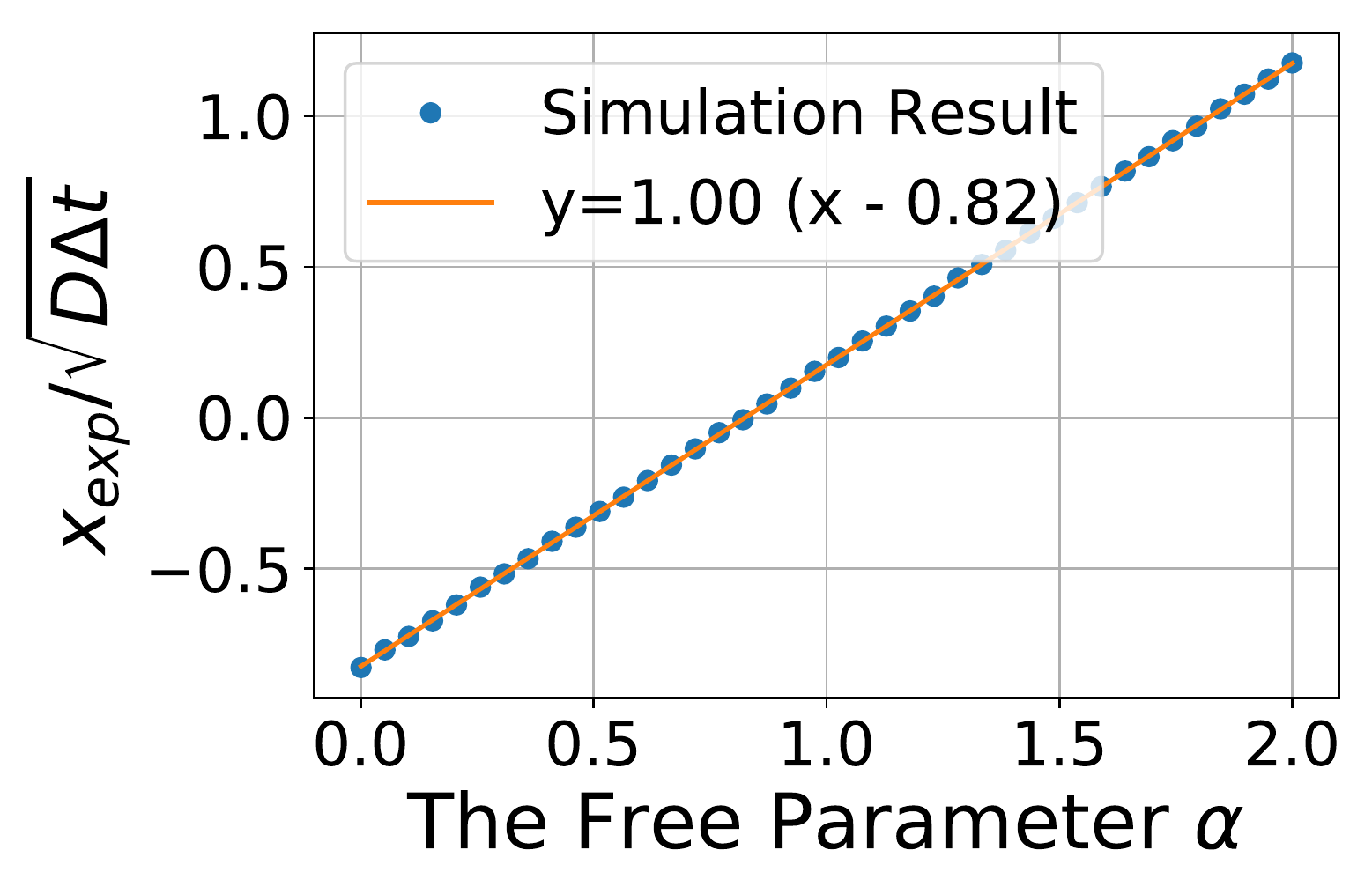}
    \caption{The linear fit describes the relationship between $\alpha$ and $x_{\mathrm{exp}}$ (in $1/\surd{D \Delta t}$) quite perfectly as expected. For this plot, a range of $L$ $(\in [30,200] \mu\rm m)$ and $D$ $(\in [80,600] \mu \rm m^2 /s)$ values are used and the mean values with error bars are plotted. The error bars are much smaller than the marker size. According to multiple trials with different parameters, the most optimum free parameter is $\alpha = 0.8235\pm 0.0005$. The standard deviation is obtained from the linear fitting performed in this plot.} \label{linear}
\end{figure}

The problems arising from the discrete time steps $\Delta t$ when diffusing molecules are encountering a boundary are illustrated in \fig{fig:main}. Besides the well-known intra-step absorption problem, we introduce the novel concept of the negative absorption index problem:
\begin{definition}[NAI Problem]
We define the absorption index  for 1D as
\begin{equation}
   AI_{1D}=x_{\mathrm{exp}}= \langle x_{\mathrm{absorbed}}-r_x \rangle
\end{equation}
where $x_{\mathrm{absorbed}}$ is the absorbed position of the molecule, $ r_x $ is the position of the receiver in 1D ($r_x=0$ unless otherwise stated) and $\langle\rangle$ denotes the average value. Finite step size simulations will always obtain $AI_{1D} < 0$, which we call the negative absorption index (NAI) problem.
\end{definition}
Note that this definition of AI is meaningful only in 1D and an analogous definition in higher dimensions requires the surface to be approximately flat on scales comparable to the step size.
For the scope of this paper, we deal only with this definition and the results that follow this intuition while leaving a more advanced treatment of the problem as future work. 

%In a conventional MC algorithm in 1-D, the molecules are absorbed if their position is less than zero, as shown in Fig. \ref{fig:MC}. This then results in a negative $x_{exp}$ value. Nonetheless, from experimental reasoning we require $x_{exp}=0$.
Consider a conventional MC algorithm, as illustrated in \fig{fig:MC}. It is clear that the NAI problem will delay the reception of molecules at the receiver's boundary and therefore deteriorate the simulation's accuracy.

%\subsection{Our Solution}
%\label{subsec_our_solution}
Conversely, a solution of the NAI problem would yield precise results for 1D simulations.
This is where the idea of locality becomes significant. If the step size is smaller than the global geometry of the diffusion channel, the absorption process can be considered local. Consequently, an individual molecule's position at distances comparable to the step size from the receiver's boundary can be approximated as independent of the global geometry of the system, hence the local problem has as its only relevant length scale the simulation step size given by $\surd(D\Delta t)$.

From this it follows immediately that formally shifting the receiver's boundary such that $r_x \to r_x + \alpha \surd(D \Delta t)$ as illustrated in \fig{fig:EGMC} solves the NAI problem if the free parameter $\alpha$ is suitably chosen.

% LFT: NOT SURE ABOUT THIS SENTENCE -- the position of the receiver is generally "unknown" to the molecules, the point is that the distribution can be considered uniform in the vicinity of the boundary I think.
%This is mainly because, if enough time passes, the molecules cannot know the position of the receiver unless they are absorbed and therefore their distribution for $x \in [0,O(\sqrt{D \Delta t})]$ will only be dependent on the step size $\Delta x \sim \sqrt{2 D \Delta t }$. 

% LFT: incorporated this into the previous paragraph
%The second realization is the fact that as the absorption of molecules is a local property which only depends on the step size, shifting the receiver position by a value $r_x \to r_x + \alpha \sqrt{D \Delta t}$ shifts the AI by the same amount, e.g. $x_{exp} \to x_{exp} + \alpha \sqrt{D \Delta t}$. This then constitutes a nice way of minimizing the NAI problem by choosing a convenient $\alpha$ which is a constant independent of $L$ and $D$. The algorithm is illustrated in Fig. \ref{fig:EGMC} and in Algorithm \ref{alg:EGMC}.

We conclude this section by illustrating the locality property of the receiver.
To this aim, we perform simulations of 1D diffusion from a source of molecules\footnote{All simulations in this work were performed with $10^5$ particles.} situated a distance $L$ from the origin, where the receiver is located. We use 100 iteration steps.
% LFT: WE NEED TO EXPLAIN HOW AND WHY WE CHOOSE THE TIMESTEP \Delta t !
In \fig{linear}, $x_{\mathrm{exp}}/\surd(D \Delta t)$ is plotted against the free parameter $\alpha$. The shape of this plot is independent of the parameters $D$ and $L$\footnote{This has been found to hold for $D \in [80, 600] \mu\rm m^2/s$ and $L \in [30,200] \mu\rm m$.}, and the interception of the linear graph with the line $x_{\mathrm{exp}} = 0$ yields the optimum choice for the free parameter $\alpha = 0.8235\pm 0.0005$ in 1D. In Appendix \ref{appendix_alpha} we show that an approximate calculation yields a similar value analytically.

\section{Application: Spherical Absorbing Receiver}
\label{sec_3D_case}
Having motivated the EG-MC algorithm from one-dimensional arguments, we now proceed to apply it to the practically more useful case of a 3D unbounded diffusion channel.
As pointed out before, it must be borne in mind that although in 1D the NAI problem was the only systematic bias, in higher dimensions the intra-step absorption problem plays a role as well.
We shall show that the EG-MC algorithm still gives results in excellent agreement with the analytic prediction for large step sizes.

\begin{figure}
    \centering
    \includegraphics[width=16cm]{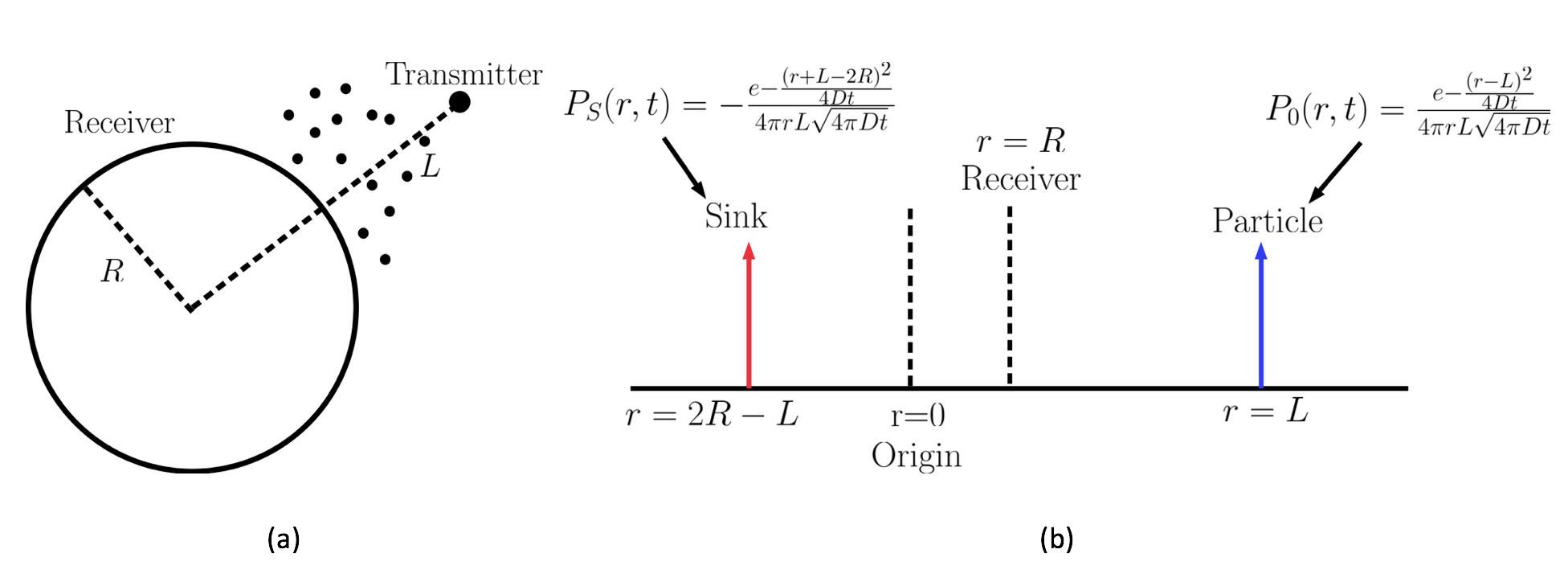}
    \caption{(a) The 3D unbounded diffusion channel consisting of a point transmitter and a spherical receiver.  (b) The 1D line, where the real space is projected onto the positive values and an imaginary space added to the negative values. In this pseudo-real (consisting of real and extended parts) space, we include a sink at $r=2R-L$ to use a more general version of method of images. } \label{fig3d}
\end{figure}

We choose our geometry such that a spherical absorbing receiver of radius $R$ is centred at the origin and receives particles from a point source situated a distance $L$ away from the origin, as illustrated in \fig{fig3d}(a).

%\subsection{Theory}
%\label{subsec_3D_theory}

In order to benchmark our simulation results, we use the analytic expressions available for this problem. They have already been derived in \cite{yilmaz2014three} in quite a tedious fashion. We simplify the calculation considerably by introducing the generalized method of images, which we illustrate in \fig{fig3d}(b). The complete derivation is given in Appendix \ref{appendix_hitting_rate}. It yields for the hitting rate, defined as the particle flux through the boundary

\begin{equation}
    n_{\mathrm{hit}}(t) =  
    \frac{R}{L} \frac{L-R}{t\sqrt{4\pi D t} } \exp(- \frac{(R-L)^2}{4 D t}),
\end{equation}

while the fraction of molecules absorbed by the receiver until time $t$ is given by

\begin{equation}
    N_{\mathrm{tot}}(t)= \int_0^t n_{\mathrm{hit}}(\tau) \diff \tau = \frac{R}{L} \text{erfc}\left[\frac{L-R}{\sqrt{4 D t}}\right].
\end{equation}

\begin{figure}
    \centering
    \includegraphics[width=18cm]{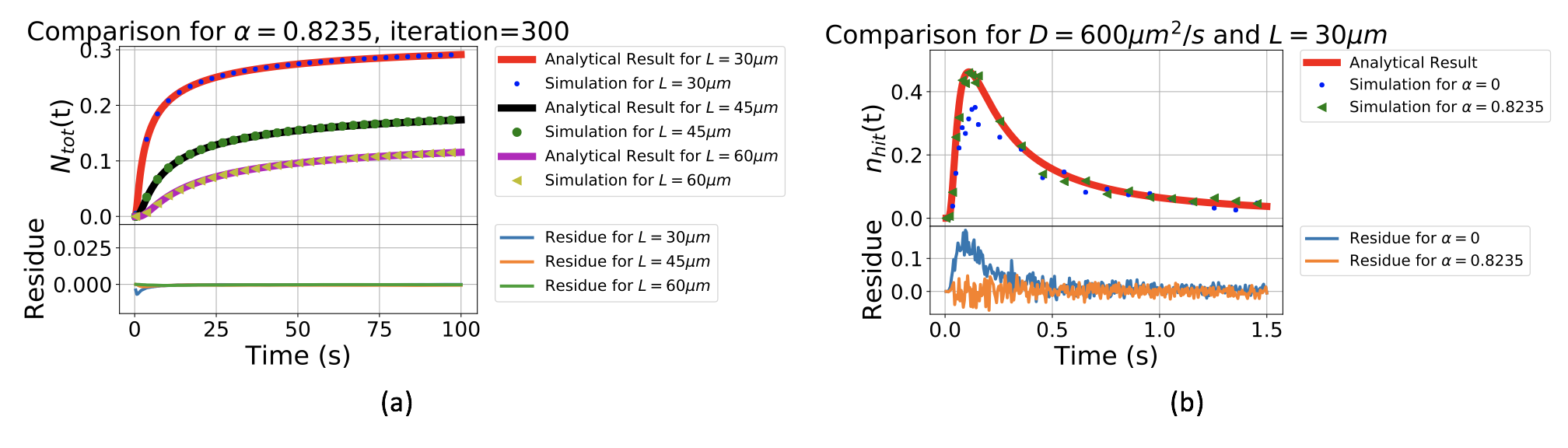}
    \caption{The proof of concept that EG-MC algorithm produces correct results for various transmitter distance $L$ (a), where $D=80\, \mu \rm m^2 /s$. The comparison of $n_{\mathrm{hit}}(t)$ obtained from both simulations with the analytical result. As can be seen, the EG-MC algorithm produces a result which follows the analytical curve closely. For both cases, the iteration number is taken as 300 and the step size can be calculated as $\Delta t= t_f/\mathrm{iteration}$, where $t_f$ is the final time of the simulation.} \label{fig3d3}
\end{figure}

\begin{figure}
    \centering
    \includegraphics[width=18cm]{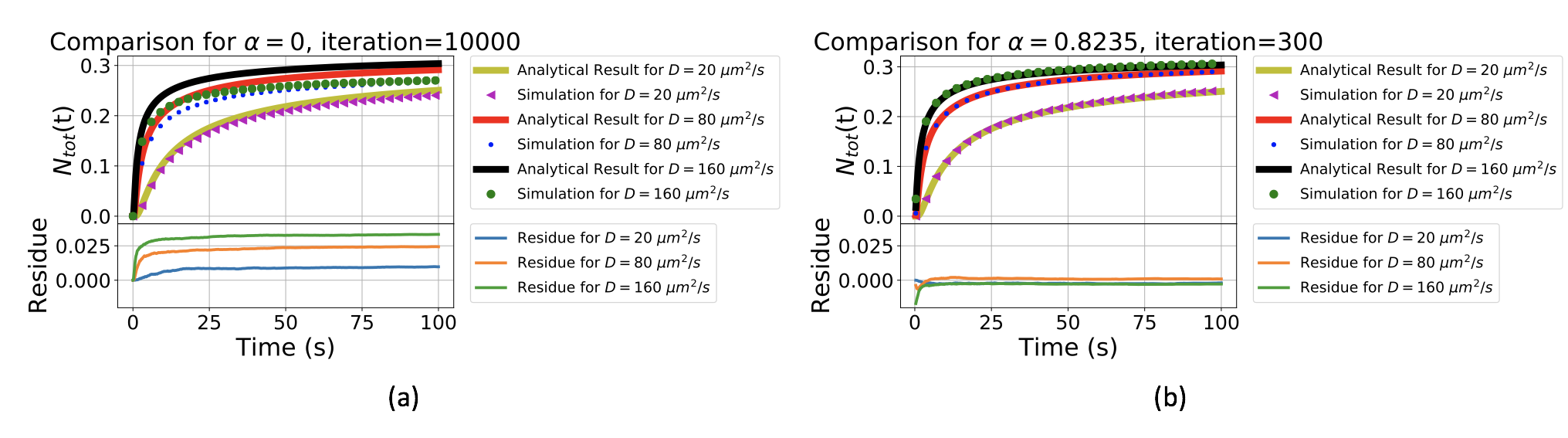}
    \caption{The comparison of Monte Carlo simulation (a) and EG-MC simulation (b) with  the analytical solution for different $D$ values and $L=30\, \rm \mu m$. For each individual case, the step size is different and therefore we note the value of total iteration. The step size can be calculated as $\Delta t= t_f/\mathrm{iteration}$, where $t_f$ is the final time of the simulation.} \label{fig3d2}
\end{figure}

%\subsection{The 3D algorithm}
\begin{algorithm}
\caption{Effective Geometry Monte Carlo (EG-MC)}\label{alg:EGMC}
\begin{algorithmic}[1]
\Procedure{Absorption Detection}{$x_i$ for i=1,2,3}
\State Define iteration number, initial position and simulation parameters ($D$,$L$,$\Delta t$)
\State $t_f = \Delta t \cdot \mathrm{iteration}$
\For{ i = 1:iteration}
\State Perform Diffusion Simulation 
\State $r= \sqrt{x^2+y^2+z^2}$ 
\State index=find($r<r_{x}+a \sqrt{D \Delta t}$)
\State $N_{\mathrm{abs}}$(i)= length(index)
\State (x,y,z)[index]=[] \Comment{Molecules are absorbed}
\EndFor
\State \textbf{end for}
\State \textbf{return} $N_{\mathrm{abs}}$
\EndProcedure
\end{algorithmic}
\end{algorithm}

We now proceed by applying the EG-MC algorithm (Algorithm 1) to the 3D diffusion channel described above.
The algorithm has the same structure as the Monte Carlo simulations, with the exception of line~7. Here, unlike usual Monte Carlo algorithms, we add a small thickness to the spherical receiver using the intuition we derived from the 1D toy model in the previous section. From now on, we denote the Monte Carlo simulations as the $\alpha=0$ limit of our algorithm. Moreover, as $\Delta t \to 0$, our algorithm converges to the Monte Carlo algorithm, as expected. 

For this section, we pick $\alpha=0.8235$ for illustrative purposes (and shall explore the constraints on the parameter in \secti{sec_error_analysis}). A comparison between simulation and analytical results is given in Figs. \ref{fig3d3} and \ref{fig3d2}. As can be seen from the figures, EG-MC outperforms the Monte Carlo algorithm significantly. Moreover, our algorithm gives robust results for far larger step sizes, hence decreasing the computational cost extremely. For example, in \fig{fig3d2}, the EG-MC algorithm converges in less than 300 iterations (we find that 20-30 iterations are the empirical cut-off for these cases where EG-MC performs quite well) while the Monte Carlo algorithm does not converge even for 10000 iterations.

Before moving on to the error analysis, we shall first define some short-hand notations. From now on, we denote the discretized values by a square bracket, e.g. $N[\cdot]$. Furthermore, we denote the fraction of particles absorbed between times $t_i$ and $t_i+\Delta t$ as
\begin{equation}
    \Delta N_{\mathrm{tot}}[t_i] = \int_{t_i}^{t_i+\Delta t} n_{\mathrm{hit}}(\tau) \diff \tau.
\end{equation}

\begin{SCfigure}
    \centering
    \caption{Dependence of ISDCD as defined in \equ{def_ISDCD} on $\alpha$ for $L = 35\, \mu \rm m$, $R = 10\, \mu m$, $D = 80\,\mu \rm m^2/s$. The simulation was run with 100 iterations for each choice of $\alpha$. Errorbars were obtained by running the simulation 20 times for each $\alpha$ and taking $\sigma_{\mathrm{ISDCD}} = \surd(\mathrm{Var}(\mathrm{ISDCD})/20)$. The fit is a simple parabola.}
    \label{fig_ISDCD}
    \includegraphics[width = 12cm]{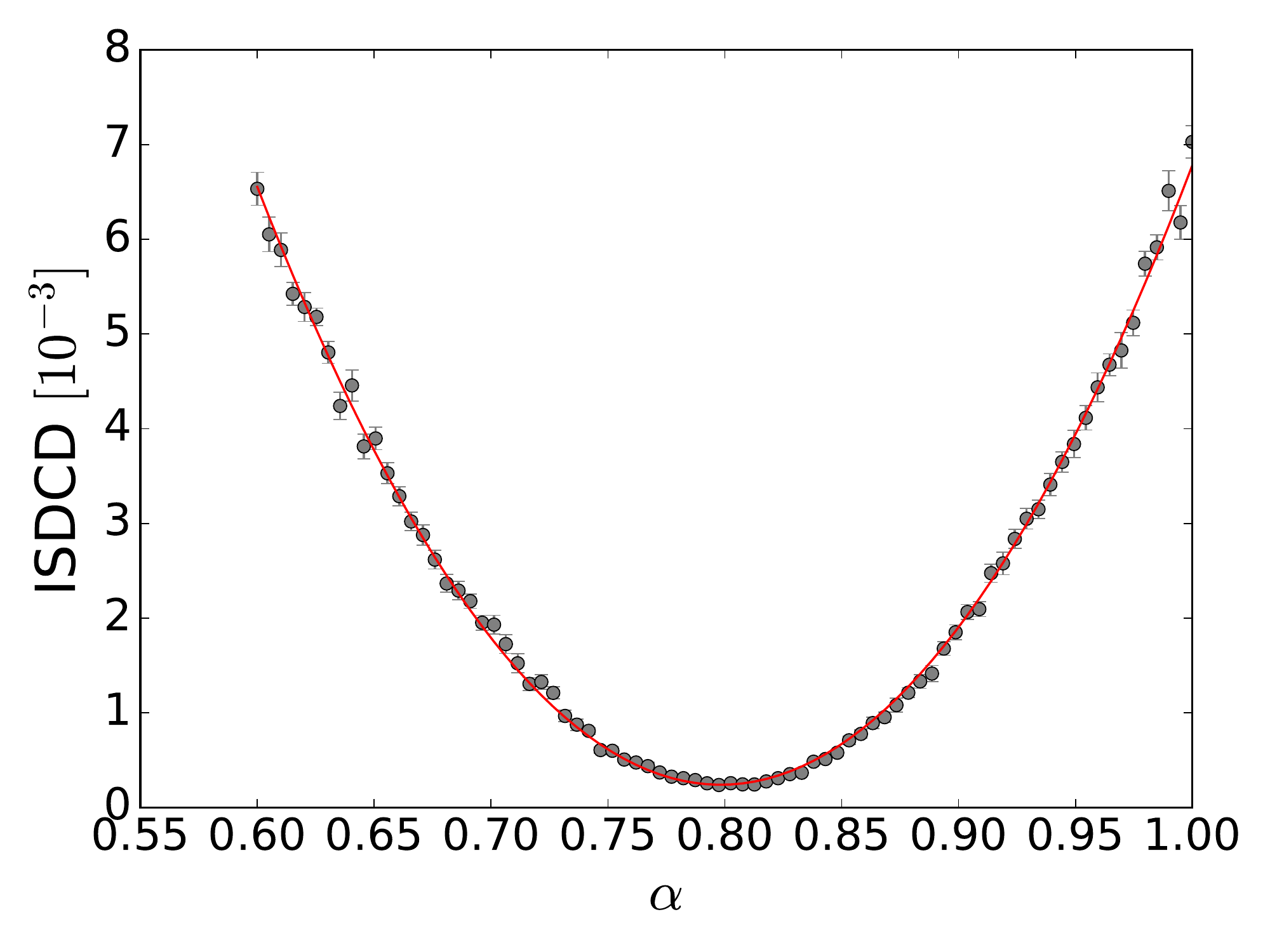}
\end{SCfigure}

\section{Error Analysis}
\label{sec_error_analysis}

In this section we discuss the uncertainty on the parameter $\alpha$ introduced above.
To this aim, we compare the simulation result for a given $\alpha$ with the analytic result.

First, we define the integrated squared difference in the cumulative distribution (ISDCD) as

\begin{equation} 
    \mathrm{ISDCD} = \sum_i (N_{\mathrm{tot}}^{\mathrm{sim}}[t_i] - N_{\mathrm{tot}}^{\mathrm{anl}}[t_i])^2,
    \label{def_ISDCD}
\end{equation}
where the sum runs over all data points in a given simulation, and ``sim" and ``anl" denote the simulated and analytical results respectively.
We choose this quantity as our primary measure for the accuracy of a given simulation for the following reasons: The cumulative distribution is a better measure than the bare probability density function (hit rate), since it mitigates the random noise in the simulation. Instead of comparing the cumulative distribution at a certain point in time, we use the sum, because it is less susceptible to random fluctuations, which are of particular importance in the steep increase of the cumulative distribution for small times.

In \fig{fig_ISDCD} we show the dependence of the ISDCD on the choice of the parameter $\alpha$. It can clearly be seen that the ISDCD has a well defined minimum around $\alpha \sim 0.8$, and follows a quadratic dependence as confirmed by the fit.

Utilizing this behaviour, we can estimate the optimum value of $\alpha$ for different choices of simulation parameters $(R, L, D)$ and different iteration numbers. This is easily done by retrieving the ISDCD, fitting it with a parabola and retrieving the value $\alpha$ that minimizes the ISDCD. A measure for the error on $\alpha$ is obtained by repeating this procedure (20 times) and taking the standard deviation of the results.

It may be argued that the better known reduced chi-squared $\chi^2_{\mathrm{red}}$ of the simulated hitting rate with respect to the analytic hitting rate is a better estimator for the goodness of the simulation, since it directly describes the simulated data. However, as argued before, this quantity is very susceptible to random fluctuations. Nonetheless, we demonstrate that the optimal choice of $\alpha$ from the ISDCD corresponds to expected values of $\chi^2_{\mathrm{red}}$. We define

\begin{equation}
    \chi^2_{\mathrm{red}} = \frac{1}{\mathrm{iteration} - 1}\sum_i \left(\frac{\Delta N_{\mathrm{tot}}^{\mathrm{sim}}[t_i] - \Delta N_{\mathrm{tot}}^{\mathrm{anl}}[t_i]}{\sigma[t_i]}\right)^{\!\!2},
    \label{def_chisquared}
\end{equation}
where we take $\Delta N_{\mathrm{tot}}$ as the bare count rate in our simulation (i.e. it is integer valued and not normalized as in the previous sections).

\begin{SCfigure}
    \centering
    \caption{Standard deviation of the hitting rate taken from 30 simulations in dependence on time. The simulations were performed with $L = 35\,\mu\rm m$, $R = 10\,\mu\rm m$, $D = 80\,\mu\rm m^2/s$, $N = 10^3$ iterations. We computed the error bars approximating the hitting rate as normally distributed.\protect\footnotemark\ The red line is the theoretical expectation for Poisson noise $\surd(\Delta N_{\mathrm{tot}}^{\mathrm{anl}}[t_i])$. For clarity, only every fifth datapoint is shown.}
    \label{fig_Poisson}
    \includegraphics[width = 12cm]{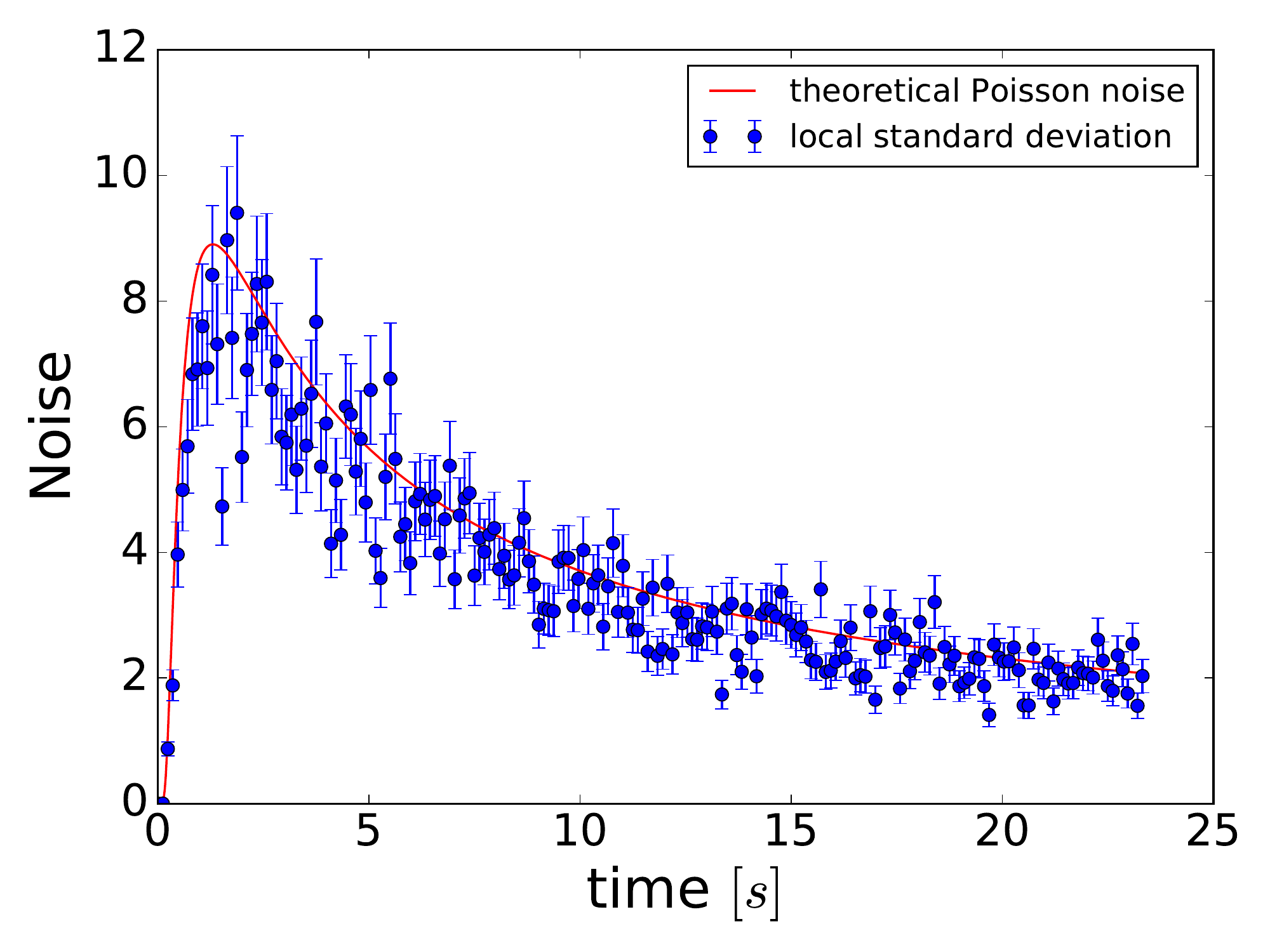}
\end{SCfigure}

We expect the random error $\sigma[t_i]$ to be close to Poisson noise.\footnote{There is a small correction due to correlations between different time steps, as a high absorption rate in one time step in general leads to a lower absorption rate in the consecutive step. However, this effect is found to be very small.} We confirm this by running the simulation repeatedly, retrieving for each datapoint the difference to the analytic result, and taking the standard deviation. The result is shown in \fig{fig_Poisson}, and confirms that we can set $\sigma^2[t_i] = \Delta N_{\mathrm{tot}}^{\mathrm{anl}}[t_i]$. This is in agreement with the various models proposed in the literature \cite{poissonnoise,poissonnoise-2}.
\setcounter{footnote}{2}
    \footnotetext{According to the formula
    $\mathrm{Var}(\sigma) = \sigma^2 \{1 - 2\Gamma^2(N/2)/[(N-1)\Gamma^2((N-1)/2)]\}$.}
\begin{SCfigure}
    \centering
    \caption{Blue datapoints are the optimal value of $\alpha$ for different input parameters, estimated from the ISDCD as described in the text. The grey triangles indicate the value of $\chi^2_{\mathrm{red}}$ obtained with this choice of $\alpha$. We also list the values of $\chi^2_{\mathrm{red}}$ for a simulation in which $\alpha = 0$ at the bottom of the plot. The average $\langle\alpha\rangle$ is weighted with the square of the individual standard deviations. The values $(R, L, D, \mathrm{iter})$ are given in units of $(\mu \rm m, \mu \rm m, \mu \rm m^2/s, 10^2)$. }
    \label{fig_alpha_with_errorbars}
    \includegraphics[width = 12 cm]{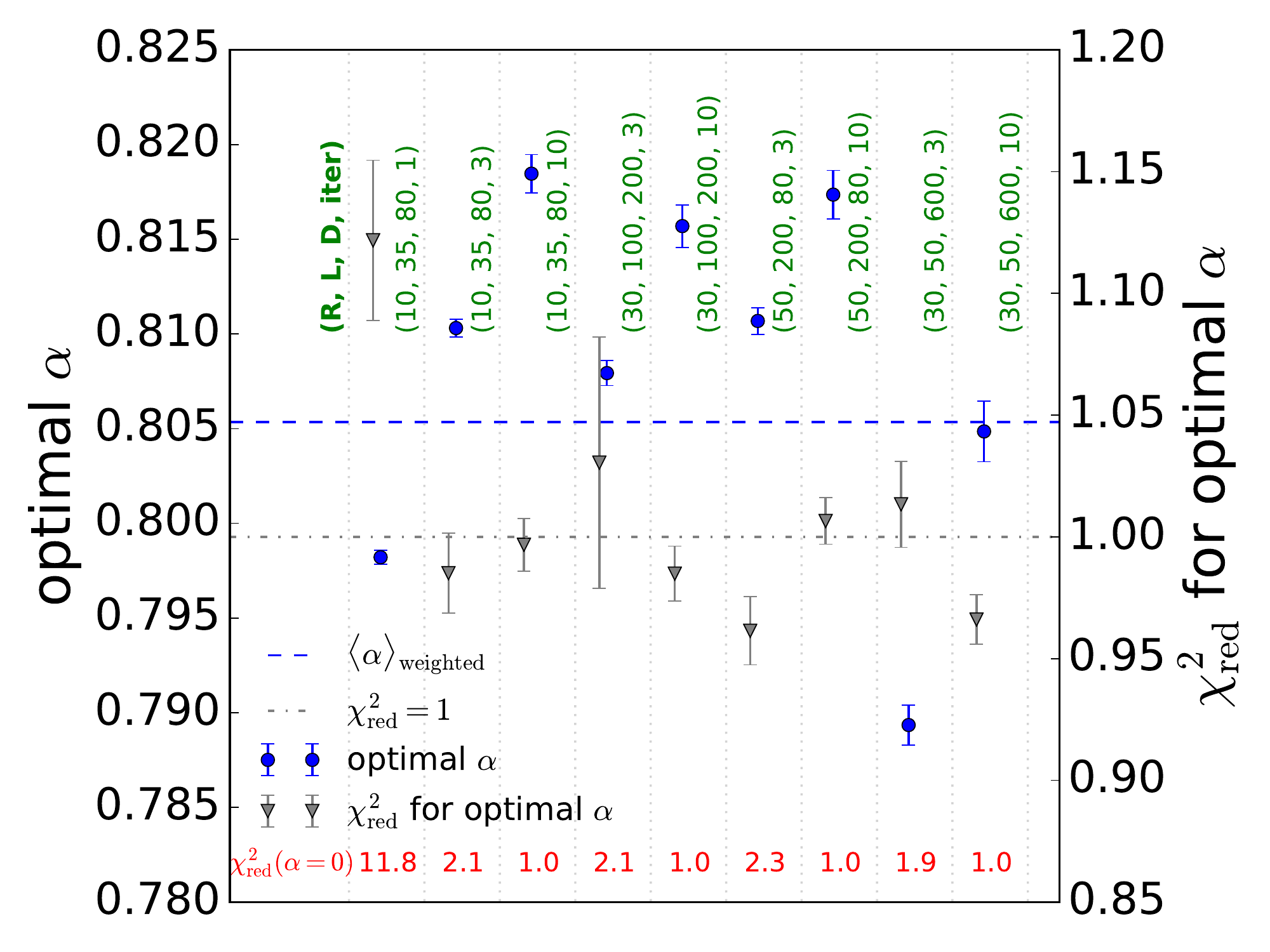}
\end{SCfigure}

In \fig{fig_alpha_with_errorbars} we show the results for the optimum value of $\alpha$ as estimated from the ISDCD, together with the corresponding $\chi^2_{\mathrm{red}}$ values from $\Delta N_{\mathrm{tot}}[t_i]$. 

We first discuss the behaviour of the optimal parameter $\alpha_O$ for different simulation inputs. It is clear that $\alpha_O$ varies only very slightly, although the simulations were performed with very different input parameters. This confirms our expectation that to a good approximation we can treat $\alpha_O$ as a constant. However, the deviations from the weighted mean are larger than expected from the errorbars. This is mainly due to a systematic effect at low iteration numbers, as for those the local problem at the receiver's boundary begins to loose its one-dimensional character which we invoked to motivate the EG-MC method.
% LFT: not sure about the next sentence, but intuitively I'd say that's the explanation
In general, we find the optimum $\alpha$ to be slightly lower in the 3D case in comparison to the 1D case, which we attribute to the fact that the $\alpha$-correction increases the total number of particles received integrated over all time (according to $R/L$), while in 1D all particles are absorbed.

Next we discuss the behavior of $\chi^2_{\mathrm{red}}$. It is well known that an optimal fit is characterized by $\chi^2_{\mathrm{red}} = 1$, and it can be seen that this value is obtained -- within statistical errors -- for each simulation shown. We also list the values of $\chi^2_{\mathrm{red}}$ for identical simulation parameters if $\alpha$ is set to 0. It is clear from these values that the EG-MC method outperforms a naive implementation by a large margin for iteration numbers of order $10^2$, while for larger iteration numbers the chi-squared is not a good measure for the goodness of fit due to the large standard deviation stemming from the Poisson noise. 

\section{Range of Applicability}

\begin{SCfigure}
    \caption{The relative inaccuracy between the Monte Carlo simulations and the EG-MC simulations for various step sizes. The iteration number is chosen such that either the final time is $6\times$ the peak value of $n_{\mathrm{hit}}(t)$ or at least 100 iterations to allow more accurate comparison between different cases. EG-MC algorithm clearly outperforms MC simulations for various step sizes. More importantly, EG-MC algorithm results in negligible error as long as $\surd(2 D \Delta t)\leq L-R$, i.e. the locality assumption is valid. } \label{fig:inac}
    \includegraphics[width=10cm]{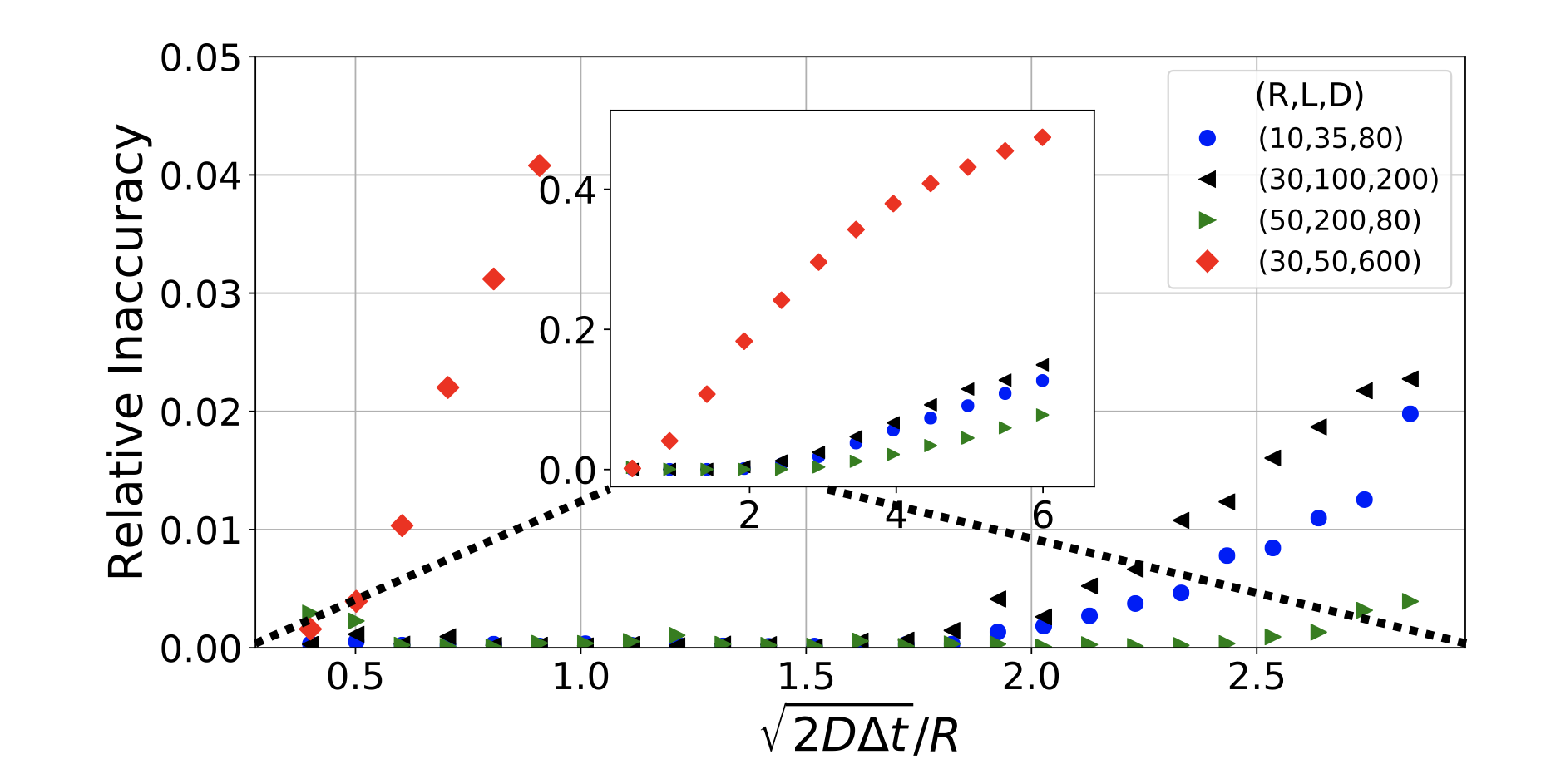} 
\end{SCfigure}

The last topic we shall address is the range of applicability of the EG-MC algorithm. As we have already discussed, the principle of locality dictates that $\surd(2 D \Delta t) \ll O(L)$. In other words, the step size should be small enough such that the molecules are not absorbed after only a few steps. To illustrate this concept, we define a comparison metric called relative inaccuracy as follows
\begin{equation}
    \text{Relative inaccuracy} =  \frac{\rm ISDCD(\alpha=0.8235)}{\rm ISDCD(\alpha=0)}.
\end{equation}
Here ISDCD($\alpha$) is the integrated squared difference in the cumulative distribution defined in \equ{def_ISDCD}, wich we evaluate for the EG-MC algorithm and compare it with a simple Monte Carlo implementation. In \fig{fig:inac}, we show the relative inaccuracy in dependence on step size for various choices of $(R, L, D)$. It shows that our algorithm performs very well even for step sizes larger than the diameter of the receiver provided that a cutoff value related to the locality condition is not exceeded. In our calculations, we find that the cutoff condition can be given more precisely as
\begin{equation}
    \sqrt{2 D \Delta t} \lesssim L-R.
\end{equation}

We can clearly see in \fig{fig:inac} that this cutoff value is observed for the four different cases we present in this paper.

\section{Conclusion}
\label{sec_conclusion}

In this paper, we have shown that an effective geometry approach can lead to significantly faster and more reliable simulations of diffusion in molecular communication. As the complete method solely consists of including a small correction -- $\alpha \surd(D \Delta t)$ -- to the receiver radius, it decreases the computational cost significantly.
% LFT: are we sure about the many orders of magnitude? I'd say this might be true for some choices of input parameters, but usually it's a factor of order 10 I believe.
In fact, we have shown, in Fig. \ref{fig3d2}, that while EG-MC converges for iteration numbers $\sim 300$, the Monte Carlo simulation does not converge to the analytical solution even for $\sim 10000$ in the case of a spherical receiver. From this perspective, our method outperforms Monte Carlo simulations by more than two orders of magnitude in terms of computation time.

We also note that the possibility of choosing a relatively large step size mitigates the problem of Poisson noise considerably, which implies that the simulation can be run with a smaller number of particles if identical noise levels for the hitting rate are to be obtained.

We believe that the EG-MC algorithm we present in this paper is just a first step towards a deeper understanding of effective geometries in the optimization of Monte Carlo type simulations.
Similar algorithms can be implemented in different fields using Monte Carlo simulations where computational cost is of primary importance.
As future work, we plan to generalize our findings to higher dimensional spaces while applying EG-MC algorithm for complex systems consisting of multiple receivers/transmitters with arbitrary shapes. Through the computational efficiency this method introduces, many systems deemed too complex to be simulated in reasonable time intervals should be considered for further research directions.

\section*{Acknowledgements}
Research at the Perimeter Institute is supported by the Government of Canada through the Department of Innovation, Science and Economic Development Canada, and by the Province of Ontario through the Ministry of Research and Innovation. LT acknowledges support by the Studienstiftung des Deutschen Volkes. FD and LT thank Matija Medvidovic for suggestions improving the code framework and Lauren Hayward Sierens for suggestions on the manuscript and error analysis. FD and BCA would like to thank Prof. Ali Emre Pusane and Prof. Tuna Tugcu for intellectually stimulating discussions on the molecular communication application of the EG-MC algorithm.

\bibliographystyle{IEEEtran}
%\balance % BALANCING THE LAST PAGE

\appendix

\section{Derivation of the Hitting Rate}
\label{appendix_hitting_rate}

%%%%
In this appendix, we derive the analytic expression for the hitting rate $n_{\mathrm{hit}}(t)$ and the integrated fraction of particles absorbed $N_{\mathrm{tot}}(t)$ in the geometry described in \secti{sec_3D_case}.

The impulse response for this diffusion channel has already been derived in \cite{yilmaz2014three}. Here, we shall re-derive their findings while motivating the solution by symmetry arguments rather than lengthy calculations. 

The diffusion of molecules in 3D is described by Fick's Law
\begin{equation}
    \grad^2 P(r,t) = D \partial_t P(r,t)
\end{equation}
where $\grad^2$ is the Laplacian operator, $P(r,t)$ is the probability distribution function of a diffusing molecule, $D$ the diffusion coefficient defined in Eq. (\ref{brown}) and $\partial_t$ is the partial derivative with respect to $t$. 

The receiver absorbs all the molecules upon incident. Moreover, the molecules are initially localized at the position $(0,0,L)$. These two conditions specify a unique solution for the diffusion equation. Nonetheless, before proceeding, we shall exploit the SO(3) symmetry. The absorption angle of the molecules is dependent on the initial position of the transmitter, whereas the absorption rate is not. As we are only interested in the absorption rate, we can solve an equivalent but easier initial value problem in which the transmitter position is symmetrised over all angles. Then, the boundary conditions become
\begin{subequations} \label{boundary}
\begin{align}
    P(r,0)&=\frac{\delta(r-L)}{4 \pi r^2}, \label{boundary1} \\
    P(R,t)&=0. \label{boundary2}
\end{align}
\end{subequations}

The calculation presented in \cite{yilmaz2014three} derives its substantial length from the boundary condition in Eq. (\ref{boundary2}). To avoid this, let us first write down the solution to the diffusion equation subject only to Eq. (\ref{boundary1}), which is describing a freely diffusing particle \cite{yilmaz2014three}
\begin{equation}
    P_0(r,t)= \frac{\exp(-\frac{(r-L)^2}{4 D t})}{4 \pi r L \sqrt{4 \pi D t}}.
\end{equation}
One important realization regarding this solution is that $L$ is a free parameter, i.e. the diffusion equation is satisfied regardless of its choice, while $L$ ensures that the boundary condition \equ{boundary1} holds.

As the system is symmetric around all angles, we can project it onto a line stretching from $0$ to $\infty$. This transformation maps $4 \pi r'^2 \diff r'$ points onto $\diff r$ points where $r \in [0,\infty)$ is the parameter describing the real line. Usually this transformation is irreversible as it is not one-to-one. Nonetheless, by the SO(3) symmetry, no information is lost during the transformation. With this in mind, we can project the complete 3D space into an incomplete 1D real space $r\in[0,\infty)$ with $r=0$ describing the origin of the original space. Then, we complete this space by adding the negative values for $r$, as shown in \fig{fig3d}(b), which do not have any physical meaning for the 3D space but are useful in solving the diffusion equation. As a final step, we include a sink at $r=2R-L$ with
\begin{equation}
    P_S(r,t) = - \frac{\exp(-\frac{(r+L-2R)^2}{4 D t})}{4 \pi r L \sqrt{4 \pi D t}},
\end{equation}
while recalling that the diffusion space is $r\geq R$. From the uniqueness theorem it follows that $P(r,t)=P_0(r,t)+ P_S(r,t)$ is the desired solution, as it satisfies the boundary conditions and the diffusion equation. Now, we can transform back to 3D space, where both $P_S(r,t)$ and $P_0(r,t)$ remain invariant due to the SO(3) symmetry.
This symmetry-motivated method is a more general version of the method of images which is used commonly in the literature.
Hence, we obtain the probability distribution function for the molecule as 
\begin{equation}
    P(r,t)= \frac{\exp(-\frac{(r-L)^2}{4 D t})-\exp(-\frac{(r+L-2R)^2}{4 D t})}{4 \pi r L \sqrt{4 \pi D t}}.
\end{equation}
Then the hitting rate, defined as the particle flux through the boundary, is given by
\begin{equation}
    n_{\mathrm{hit}}(t) = 4 \pi R^2 D\,\partial_r P(r,t)|_{r=R} = \frac{R}{L} \frac{L-R}{t\sqrt{4\pi D t} } \exp(- \frac{(R-L)^2}{4 D t}).
\end{equation}
The fraction of molecules absorbed by the receiver until time $t$ follows as
\begin{equation}
    N_{\mathrm{tot}}(t)= \int_0^t n_{\mathrm{hit}}(\tau) \diff \tau = \frac{R}{L} \text{erfc}\left[\frac{L-R}{\sqrt{4 D t}}\right].
\end{equation}
%%%

\section{Analytical Approximation for $\alpha$}
\label{appendix_alpha}

In this appendix, we are deriving an approximate expression for $\alpha$ in the 1D case.
We choose our geometry such that the receiver is located at $x = 0$, and the effective receiver at $\alpha'$.
Let us consider a single time step in the simulation. At the beginning of the step, a particle is located at $x_0 > \alpha'$. We first find the expected absorption position $\langle x_1(x_0)\rangle$ for this individual particle under the assumption that it is absorbed. We choose our units of length such that the probability density function for the location of this particle after the time step is given by
\begin{equation}
    P(x, x_0) =N \mathrm{e}^{-(x-x_0)^2},
\end{equation}
where we omit normalization factors and lump them in $N$ because they are of no importance for the calculation and would only serve to clutter the expressions.

Hence we find that
\begin{equation}
    \langle x_1(x_0) \rangle = \frac{\int_{-\infty}^{\alpha'}\diff x\,x P(x, x_0) }{\int_{-\infty}^{\alpha'}\diff x\,P(x, x_0)}.
\end{equation}

The probability of absorption is given by

\begin{equation}
    P_{\mathrm{abs}}(x_0) = \int_{-\infty}^{\alpha'} \diff x\,P(x, x_0).
\end{equation}

Denoting the instantaneous distribution of particles by $\rho(x_0)$, the expected absorption position can then be given by using the probability distribution ($P_{\mathrm{abs}}(x_0)\rho(x_0) \diff x_0$) as
\begin{equation}
    \langle x_{\mathrm{abs}}\rangle =\frac{ \int_{\alpha'}^{\infty}\diff x_0\,\langle x_1(x_0)\rangle P_{\mathrm{abs}}(x_0) \rho(x_0)}{\int_{\alpha'}^\infty \diff x_0\,P_{\mathrm{abs}}(x_0) \rho(x_0)},
\end{equation}

We now want to find $\alpha'$ such that the expected absorption position (for all particles present at $x > \alpha'$) equals zero. It follows that

\begin{equation}
    0 = \int_{\alpha'}^{\infty}\diff x_0\,\langle x_1(x_0)\rangle P_{\mathrm{abs}}(x_0) \rho(x_0),
\end{equation}

We now invoke the approximation that the distribution $\rho(x_0)$ is uniform. This approximation will yield a relatively precise result for $\alpha$ because only particles relatively close to the receiver have an appreciable chance of absorption, which justifies a zeroth order approximation.

Inserting the expressions given before, we find

\begin{equation}
    0 = \int_{\alpha'}^{\infty}\diff x_0\, \int_{-\infty}^{\alpha'}\diff x\,x \mathrm{e}^{-(x-x_0)^2}.
\end{equation}
Performing the change of variables $\chi \equiv x-x_0$ in the second integral,  we find
\begin{equation}
    0 = \int_{\alpha'}^{\infty} \diff x_0\,
    \int_{-\infty}^{\alpha'-x_0}\diff\chi\,\left(\chi\mathrm{e}^{-\chi^2} + x_0\mathrm{e}^{-\chi^2}\right) =
    \int_{\alpha'}^{\infty}\diff x_0\,\left(
    \frac{1}{2}\mathrm{e}^{-(x_0 - \alpha')^2}
    + x_0 \frac{\surd\pi}{2}\mathrm{erfc}(x_0 - \alpha')
    \right).
\end{equation}
We simplify the final integration using $\xi \equiv x_0 - \alpha'$, so that
\begin{equation}
    0 = \int_0^{\infty}\diff\xi\,\left(
    \mathrm{e}^{-\xi^2} +
    \xi\surd\pi\mathrm{erfc}(\xi) +
    \alpha'\surd\pi\mathrm{erfc}(\xi)
    \right),
\end{equation}
from which it follows that
\begin{equation}
    \alpha' = \frac{\frac{1}{2} - \int_0^{\infty}\diff\xi\,\xi\mathrm{erfc}(\xi)}{\int_0^{\infty}\diff\xi\,\mathrm{erfc}(\xi)} = \frac{\sqrt \pi}{4}.
\end{equation}

Returning to the units of length used in the main body of the text, we find $\alpha = 2\alpha' = \sqrt \pi/2 \simeq 0.89$.
Hence, we see that using the approximation of a uniform $\rho(x_0)$, a result within 7\% of the empirically found value can be obtained.

\end{document}